# Large heat flux in electrocaloric multilayer capacitors

Authors

Romain Faye[*], Hervé Strozyk[*], Brahim Dkhil[+*], Emmanuel Defay[*]

[*] Materials Research & Technology Department, Luxembourg Institute of Science and Technology (LIST), 41 Rue du Brill, L-4422 Belvaux, Luxembourg

[+] Laboratoire Structures, Propriétés et Modélisation des Solides (SPMS), CentraleSupélec, CNRS UMR8580, Université Paris-Saclay, 92290 Châtenay-Malabry, France

Abstract

Multi Layer Capacitors (MLCs) are considered as the most promising refrigerant elements to design and develop electrocaloric cooling devices. Recently, the heat transfer of these MLCs has been considered. However, the heat exchange with the surrounding environment has been poorly, if not, addressed. In this work, we measure by infrared thermography the temperature change versus time in four different heat exchange configurations. Depending on the configurations, Newtonian and non-Newtonian regimes with their corresponding Biot number are determined allowing to provide useful thermal characteristics. Indeed, in case of large area thermal pad contacts, heat transfer coefficients up to 3400 $W.m^{-2}.K^{-1}$ are obtained showing that the standard (non-optimised) MLCs already reach the needs for designing efficient prototypes. We also determine the ideal Brayton cooling power in case of thick wires contact which varies between 3.4 mW and 9.8 mW for operating frequencies varying from 0.25Hz to 1Hz. While only heat conduction is considered here, our work provides some design rules for improving heat exchanges in future devices.

Introduction

Electrocaloric (EC) materials display a reversible adiabatic change of temperature when an external electric field is applied and withdrawn. Although the effect has been discovered in ceramics since the thirties, it is only recently that EC-based prototypes have been considered with only a few examples published ( (Sinyavskii, 1995), (Jia & Jua, 2012) , (Gu, et al., 2013) , (Crossley, et al., 2014), (Wang, et al., 2015), (Plaznik, et al., 2015), (Blumenthal, et al., 2016) , (Sette, et al., 2016)). One of the main issues is to extract efficiently the heat from the EC material i.e. being simultaneously able to apply a strong enough electric field and keep enough matter to extract maximum heat. This is why Multi Layer Capacitors (MLCs) – a stack of EC material layers with intercalated conductive layers with high thermal conductivity – that combine the aforementioned criteria raise a strong interest in the community. As a result, a better knowledge concerning the thermal behaviour of MLCs is crucial in order to integrate them into any refrigerator prototype ( (Jia & Jua, 2012), (Wang, et al., 2015) , (Sette, et al., 2016) ) and to determine  the limits that MLCs themselves could impose to the operating conditions. While more sophisticated than bulk EC materials with one top and one bottom electrodes, MLCs suffer (are limited by) from electrocalorically inactive regions.  Indeed, in addition to the active part made of ceramic layers sandwiched between electrodes and undergoing the electric field, there is an inactive part composed of both metal electrodes and terminals and ceramic areas not experiencing the field. The latter areas are located close to the terminals and at the edge of the MLC. Understanding the thermal behaviour of MLCs requires therefore distinguishing between inner or proper heat conduction that corresponds to the heat transfer from the active part to the inactive one within the MLC and the external heat exchange corresponding to that between the MLC and the surrounding environment. Some predictions of the MLCs thermal behaviour have been made by either modelling the peculiar multilayers heterostructure of MLCs ( (Kar-Narayan & Mathur, 2009), (Crossley, et al., 2014)) or measuring their surface heat flux ( (Liu, et al., 2016), (Liu, et al., 2016)). Kar-Narayan *et al.* (Kar-Narayan & Mathur, 2009) and Liu *et al.* (Liu, et al., 2016) considered lumped thermal modelling to describe the heat exchange inside MLCs. In these works only the inner heat flux from active part to MLC terminals has been taken into account. By determining an equivalent thermal circuit, cooling power and relaxation time constant have been thus estimated for different MLC configurations. It has been demonstrated, as one would expect, that inner metallic electrodes play an important role in conducting the heat between EC active and inactive regions because of their good thermal conductivity. The cooling power of commercially available MLCs through their terminals is found to range between 0.3 kW.m$^{-2}$ and 1 kW.m$^{-2}$. Moreover, Crossley *et al.* (Crossley, et al., 2014) foreseen a potential enhancement of the cooling power that could reach a few tens of kW.m$^{-2}$ by optimizing materials and geometry of MLCs. However, the external heat exchange of the MLC and its near environment has not been regarded. In this paper, we therefore investigated *in situ* the heat exchange between an MLC and its surrounding environment considering the MLC in contact with metal. After the seminal work of Kar-Narayan *et al.* (Kar-Narayan, et al., 2013), we monitored using infrared (IR) camera the temperature change of the whole MLC as a function of time for different contacts to reveal the impact of external metallic solids in extracting/absorbing the heat from the MLC. Interestingly, it is found that a cooling power peaking at 13 kW.m$^{-2}$ and reaching 4.3 kW.m$^{-2}$ during the first 500 ms can be achieved using typical commercial MLCs. Although our study is limited to solid-solid thermal exchange, it experimentally confirms the ability of MLCs for exchanging heat.

Methods section

MLCs are purchased from Multicomp company. They display a capacitance value of 47μF (MULTICOMP MC1210F476Z6R3CT) and are made of doped BaTiO$_3$ and nickel inner metal electrodes. Their size is 2.5 mm wide, 2 mm thick and 3.2 mm long. The temperature change measurements are carried out with a FLIR X6580SC infrared camera that exhibits a typical temperature resolution of 20 mK. The measurements are performed at 145 Hz. In order to ensure emissivity homogeneity, the MLCs are slightly polished and black painted. Emissivity is determined by measuring with the IR camera the apparent temperature of a diffuse IR reflector (in our case a creased aluminium foil) and the measured temperature of the sample is put at the reference temperature (ASTM International, 2014). Consequently by measuring the actual temperature with a thermocouple, we have access to the surface emissivity value. In these conditions the emissivity reaches 0.96, which is taken into account in the IR camera post-treatment software. During the measurements, the MLC is suspended by wires to avoid unintended thermal exchange. 200 volts are applied at 10 mA constant current by using a Keithley 2400 sourcemeter. The temperature is measured on the whole surface of the MLC as a function of time after voltage application. The magnitude of the EC response (not shown here) is identical in both cooling and heating processes, which shows that Joule heating effect, if any, is negligible.

Results

Figure 1 shows the 4 different cases we consider in this study. The reference case (case #1) is an MLC with two 0.1 mm-thick copper electrical wires soldered to the MLC terminals. For the second case (case #2), the two wires are replaced by two thicker wires of 1.7 mm-thick. The low surface thermal resistance between the solder and the wire increases the heat exchange with the MLC. In the two other cases a 0.5 mm-thick thermal pad (Arctic thermal pad ACTPD00002A with 6 W.m$^{-1}$.K$^{-1}$ thermal conductivity – Arctic company) is used to contact the MLC sides (case #3) or terminals (case #4) with two copper blocks weighing about 10.3 g each. In cases #3 and #4, the use of thermal pad permits to compare heat exchange on ceramic side and metallic terminal side, respectively.

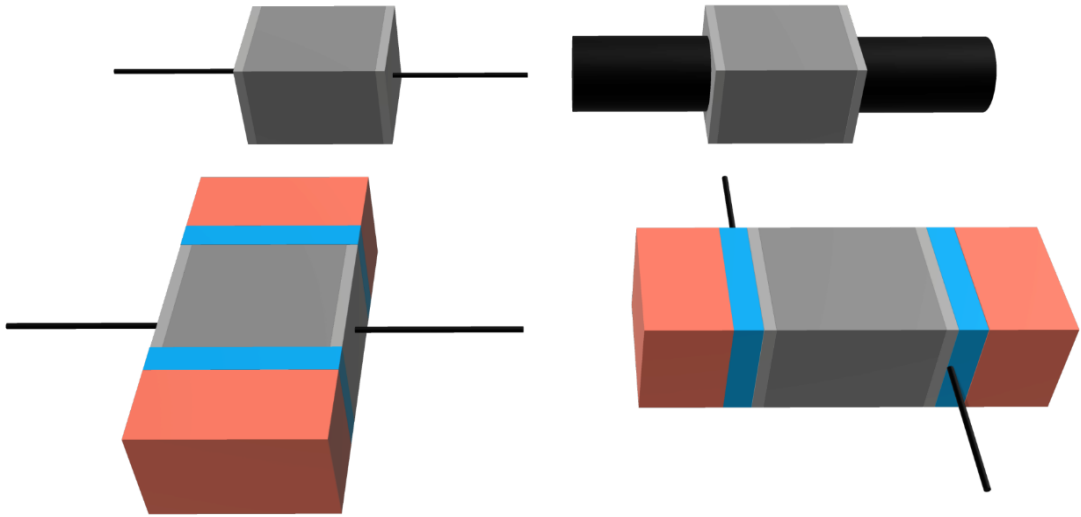

*Figure 1. Schematic view of the different investigated heat exchange configurations. Electrical wires are depicted in black, thermal pad in blue and copper in red. Case #1 - reference case: two 0.1 mm thick wires soldered to each terminal. Case #2 - two 1.7 mm thick wires soldered to each terminal. Case #3 – two 0.1 mm thick wires soldered to each terminal and thermal contact between copper and sides via 0.5mm thick thermal pad. Case #4 - two 0.1 mm thick wires soldered to each terminal and thermal contact between copper and terminals via 0.5 mm thick thermal pad.*

The amount of available heat $Q_{MLC}$ is calculated by using $Q_{MLC} = m_{MLC}.Cp_{MLC}.\Delta T_{MLC}$ where the mass of the MLC $m_{MLC}$ is of 85.6 mg and $Cp_{MLC}$ the MLC specific heat capacity measured by Differential Scanning Calorimetry is 400 J.kg$^{-1}$.K$^{-1}$ at 25°C. Therefore, monitoring the MLC average temperature $\Delta T_{MLC}$ versus time yields direct insight into how fast the heat is extracted or absorbed.

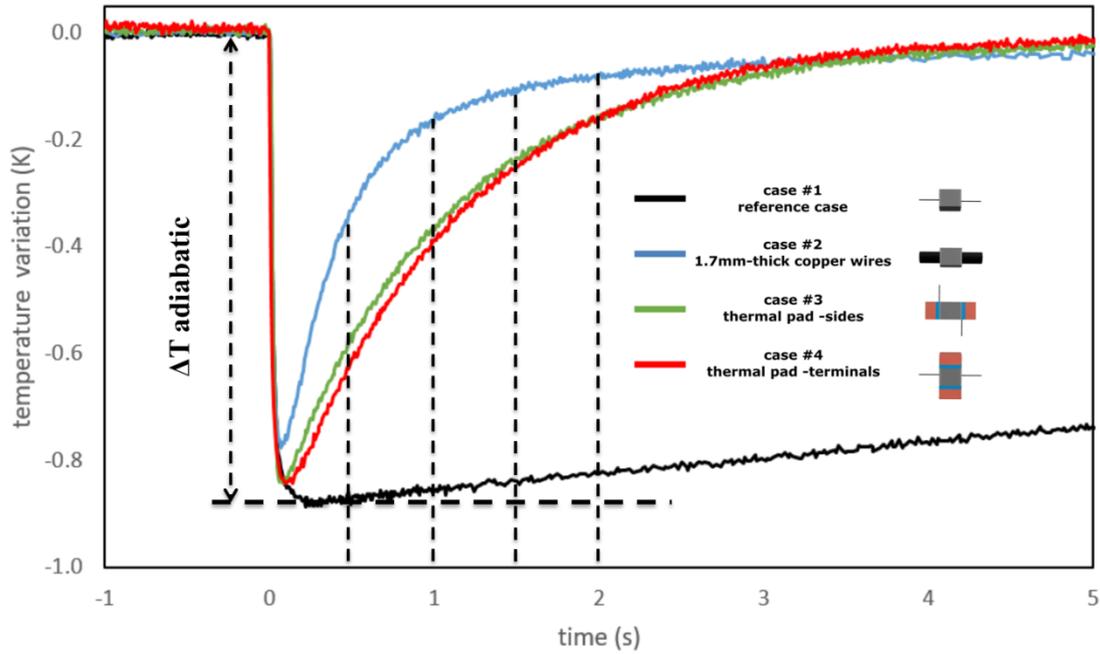

*Figure 2: Average MLC temperature change as a function of time according to the four different heat conduction configurations depicted in insert. The electric field has been removed at t = 0 s. The dashed lines are useful to explain how heat has been extracted from the temperature variation of case #1 and case #2 (cf text for details)*

Figure 2 shows the temperature variation $\Delta T_{MLC}$ just after the electric field has been removed for the four different heat exchange configurations described in Figure 1. t = 0 s corresponds to the time at which the electric field has been withdrawn. Differences in maximum amplitude of *ΔT* occur when comparing the reference case #1 with case #2 (thicker wires) and cases #3 and #4 involving thermal pads. In these cases #2, #3 and #4, the EC temperature change *ΔT* of the MLC cannot be considered as under fully adiabatic conditions. Indeed, a part of the heat starts being exchanged with the surrounding environment (mainly the metal wire or pad contact) at the moment the field is removed, which prevents MLCs from reaching the expected maximum amplitude *ΔT* as in case #1. Inner heat conduction and external heat exchange (see Figure 3) occur at the same time. In Figure 2, one can immediately see large differences in terms of heat exchange as *ΔT* dramatically varies in time depending on the configuration considered. Indeed, in case #1, during the first second, the proportion of exchanged heat reaches only 3% whereas it reaches 84% for case #2 and about 60% for both cases #3 and #4. Here, we are dealing with non-steady state heat transfers therefore, we have to distinguish between two different regimes, namely Newtonian and non-Newtonian heat transfers.

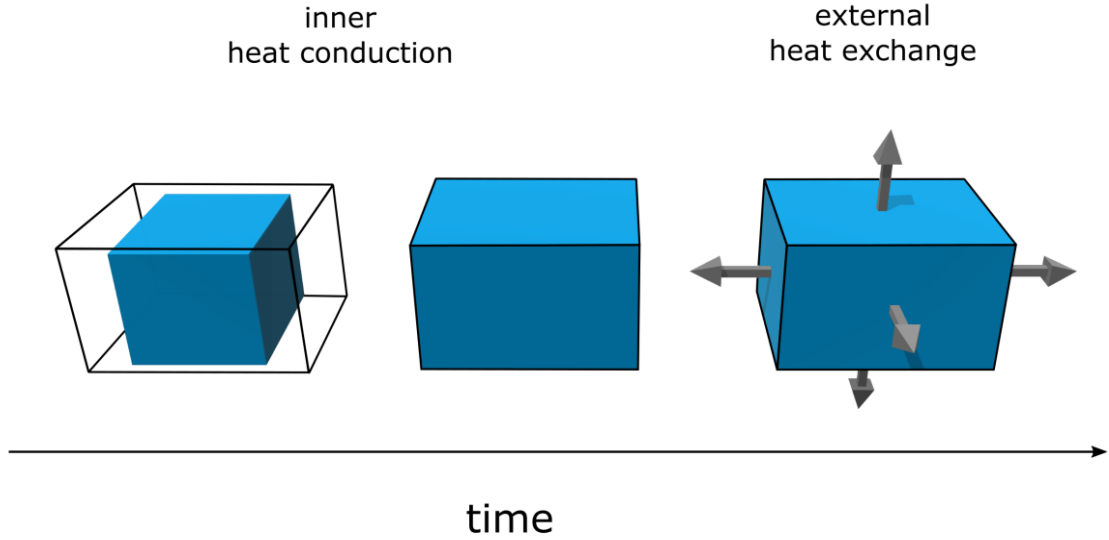

*Figure 3: Decomposition in time of the thermal exchange process just after the electric field is removed (or applied). For sake of simplification, we represented isotropic heat flow on this cartoon, though in reality the electrodes may preferentially carry heat to the terminals*

To be more specific, these two regimes of heat transfer can be distinguished through the Biot number *Bi* defined as:

$$Bi = \frac{h.L_C}{k}$$

where $h$ is the heat transfer coefficient between the body surface and the surroundings, $k$ is the thermal conductivity of the specimen, $L_C$ is the characteristic length defined as the ratio of the body volume to its surface area ($L_C = \frac{V}{A}$). *Bi* is a dimensionless quantity. It has been arbitrarily defined that Bi= 0.1 is the limit between Newtonian and non-Newtonian heat transfer. In other words, for a given temperature difference, the heat transfer is Newtonian if heat flow resulting from inner thermal conductivity is at least 10 times larger than heat flow resulting from external thermal transfer. In this Newtonian case, one can consider that there is no temperature gradient into the MLC. In case of Newtonian regime, the temperature of the body can thus be considered as uniform in all its parts during the transfer. Here, *Bi* has been estimated for the MLC in the four different cases displayed in Fig. 2 extracting *h* values from the literature (Otiaba, et al., 2011). *h* value typically varies between 0.5 and 1000 W.m$^{-2}$K$^{-1}$ in the case of free air convection (similar to case #1), between 1000 and 3300 W.m$^{-2}$K$^{-1}$ with thermal gel pads (cases #3 and #4) and is above 200 000 W.m$^{-2}$K$^{-1}$ in the solder case (case #2). $L_C$ values depend on the considered exchange surface, i.e. the entire surface - 39 mm² - in case #1, 4.5 mm² corresponding to the wire section of the 1.7 mm thick wires in case #2, 13 mm² in case #3 (gel pad on MLC sides) and 10 mm² in case# 4 (gel pad on MLC terminals). Consequently $L_C$ in the four different cases is respectively equal to 0.4 mm, 3.5 mm, 1.3 mm and 1.6 mm. Two extreme cases have been considered regarding the thermal conductivity of the MLC ($k_{MLC}$) that is made of BaTiO$_3$–based ceramic and Ni internal electrodes $k_{BaTiO3} = 6\ W.m^{-1}.K^{-1} < k_{MLC} < k_{Nickel} = 94\ W.m^{-1}.K^{-1}$. The determined *Bi* are displayed in Figure 4.

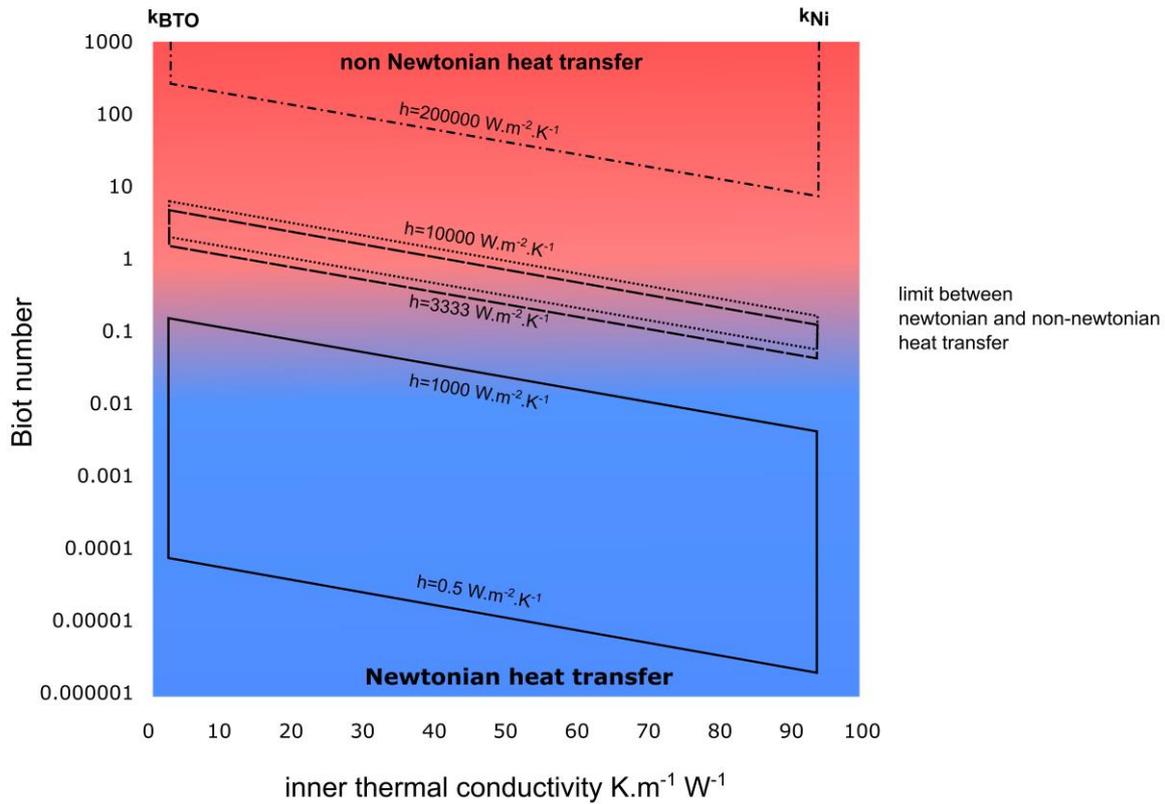

*Figure 4: Biot numbers Bi as a function of inner conductivity in the four different heat exchange cases displayed in Fig 2. Biot numbers have been estimated using a range of h values. In the bottom part of the graph (blue) the heat transfer is Newtonian, whereas it is non-Newtonian in the top part of the graph (red). The blurred interface reminds that Bi=0.1 has been defined arbitrarily and should not be considered as a sharp cut-off.*

As shown in Figure 4, for case #1 *Bi* is almost always below 0.1, whatever the inner thermal conductivity. Therefore the heat transfer is limited by the external heat exchange, outside the MLC. In case #2 (thick copper wires), the heat transfer is limited by the internal heat exchange. *Bi* is well beyond 0.1 for any value of *k*. In cases #3 and #4 (thermal pads), the range of *Bi* value is very close to the boundary between both transfer regimes. *Bi* value can thus be below or above 0.1 depending on both *k* and *h* values, though *Bi* typical value is 1, meaning that inner heat conductivity and external heat transfer are of the same order of magnitude. One of the consequences of Newtonian heat transfer is that the MLC temperature change as a function of time can be described by a decreasing exponential in time following equation (1):

$$\frac{T(t)-T_{env}}{T_i-T_{env}} = \exp\left(\frac{-h.A.t}{m.C_p}\right)$$

where $T_{env}$ is the surrounding environment temperature, *T(t)* is the MLC temperature at time *t*, $T_i$ is its initial temperature i.e. *T(t)* = $T_i$ when *t* = 0 s, m its mass, $C_p$ its specific heat capacity, A its exchange area and h is the external heat transfer coefficient. We use equation (1) to fit our experimental data in cases #1, #3 and #4 where a Newtonian regime is or could be (for #3 and #4) expected with respect to *Bi* domains found in Figure 4. The results are shown in Figure 5. Figure 5 confirms that in case #1 the heat exchange is a Newtonian one. The rather good agreement between the fit and experimental data in cases #3 and #4 shows that the heat exchange can reasonably be considered as Newtonian as well in these both cases. We can now in these 3 Newtonian-like configurations extract the heat transfer coefficient which yields to 35 W.m$^{-2}$K$^{-1}$ for case #1, 2600 W.m$^{-2}$K$^{-1}$ for case #3 and 3400 W.m$^{-2}$K$^{-1}$ for case #4. These values are comparable to values reported in the literature (Otiaba, et al., 2011).

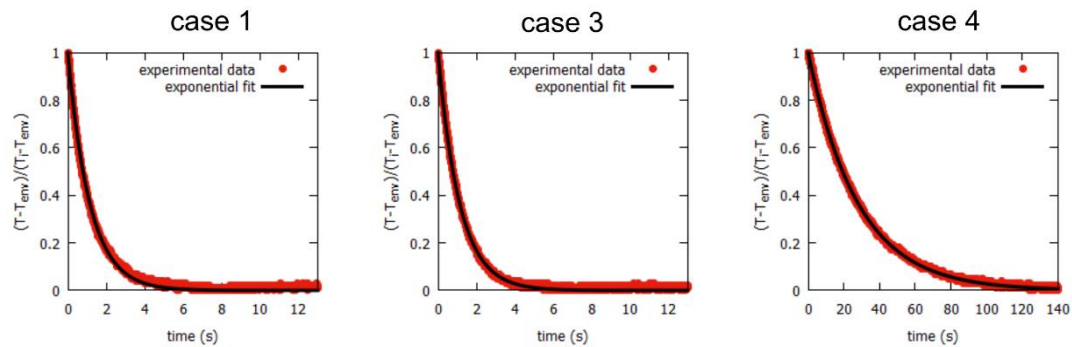

*Figure 5: (T(t)-T$_{env}$)/(T$_i$-T$_{env}$) as function of time for the 4 different cases. In red, a decreasing exponential fit for each case.*

As expected, the temperature variation in case #2 cannot be fitted with equation (1), confirming that in such situation the heat transfer is not Newtonian and is thus limited by the inner thermal conductivity and not by the external heat exchange. *h* is thus not constant but should be larger than in cases #3 and #4, i.e. larger than 3400 W.m$^{-2}$K$^{-1}$. Moreover, as in case #2 we have access to the MLC intrinsic thermal behaviour, it is possible to extract some thermal characteristics that describe the ability of this MLC to exchange heat. Eventually, the very beginning of the temperature curves versus time are linearly fitted in order to extract the maximum heat exchange power *P$_{max}$* independently from the exchange surface and the temperature difference between the sample and the surroundings. $\Delta T_{MLC}$ = -0.87 K extracted from case #1 has been used to calculate $Q_{MLC}$ because it is the only case from which it is possible to have access to adiabatic $\Delta T_{MLC}$. We considered that the same amount of heat has been created in case #2. Therefore $Q_{MLC}$ reaches 30 mJ at 200V and 25°C.The maximum value *P$_{max}$* is then obtained for case #2 and reaches 60 mW.

Discussion

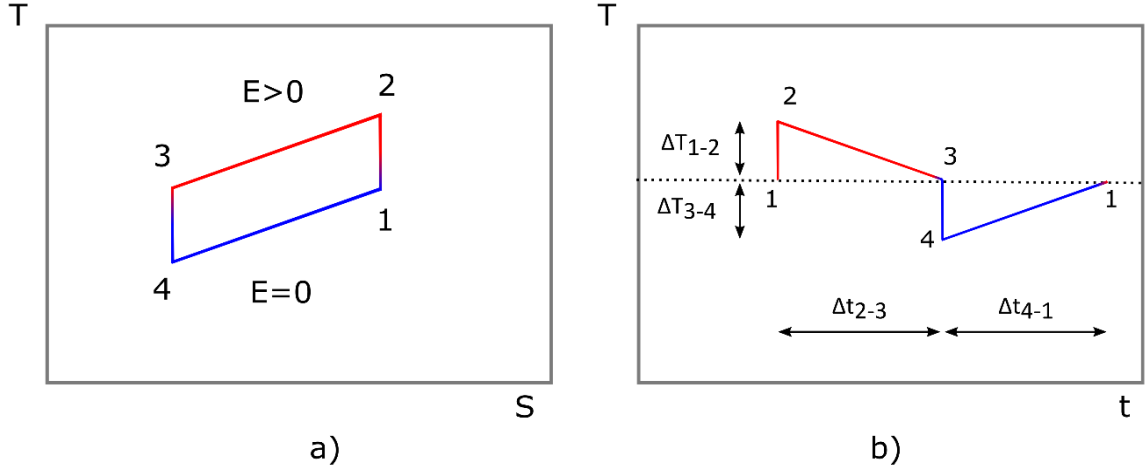

*Figure 6 -a. Schematic Brayton refrigeration cycle - b. Brayton cycle steps represented as a function of time.*

By exploiting case #2's results, we can evaluate what the MLC cooling power ($P_B$) could be for an ideal Brayton cycle at different operating frequencies ($f_o$). Let us consider the simple Brayton refrigeration cycle represented in Figure 6.a and described as follows: (1-2) the first step is an adiabatic increase of temperature as a consequence of the application of an external electric field, (2-3) followed by heat exchange between the EC material and the hot part of the refrigerator, (3-4) then electric field is removed and the temperature of the active part decreases adiabatically, (4-1) finally the material absorbs heat coming from the cold part and sets back to temperature 1. The representation of this refrigeration cycle as a function of time (Figure 6.b) shows that the two adiabatic steps are instantaneous contrary to the isofield ones.

The operating frequency can be written as $f_o = \frac{1}{\Delta t_{2-3} + \Delta t_{4-1}}$ with $\Delta t_{2-3}$ the time delay to extract heat after the application of an electric field (corresponding to step 2-3 in Figure 6) and $\Delta t_{4-1}$ the one to absorb heat after the electric field has been removed (corresponding to step 4-1 in Figure 6.b). Here, we took $\Delta t_{2-3} = \Delta t_{4-1}$. Besides, the duration of the two adiabatic steps were neglected (corresponding to step 1-2 and 3-4 in Figure 6). The quantity of heat $m.C_p.|\Delta T_{4-1}|$ exchanged during the 4-1 leg in figure 6a was extracted from Figure 2. Although adiabatic conditions were not fulfilled in case #2 because of rapid heat exchange, we considered that $|\Delta T_{4-1}|$ is $|\Delta T_{adiabatic}|$ = 0.87 K minus the temperature magnitude reached in case #2 after respectively 0.5 s, 1 s, 1.5 s and 2 s (cf dashed lines in figure 2 for details). Indeed the amount of heat due to the EC effect is very likely to be similar in both cases #1 and #2. The ideal Brayton cooling power $P_B$ at $f_o$ was subsequently defined as $P_B = \frac{m.C_p.|\Delta T_{4-1}|}{(\Delta t_{2-3} + \Delta t_{4-1})}$. The $P_B$ values obtained at 1Hz, 0.5Hz, 0.33Hz and 0.25 Hz were respectively 9.8 mW, 6.2 mW, 4.3 mW and 3.4 mW. It is worth mentioning that the obtained experimental cooling powers are comparable to the MLC cooling power reported by Kar-Narayan et al. (2.15 mW at 0.33 Hz) (Kar-Narayan & Mathur, 2009) and Crossley et al. (11.7 mW at 0.25Hz) (Crossley, et al., 2014).

The next question is how this value is aligned with prototyping needs in terms of heat exchange. An experimental heat exchange coefficient of 1000 W.m$^{-2}$.K$^{-1}$ was reported in Sinayavskii *et al.*'s seminal work in the nineties considering forced convection heat transfer conditions between ceramic plates and hexane fluid. In the same work, a modelling estimation shows that being able to increase this coefficient up to 5000 W.m$^{-2}$.K$^{-1}$ is required to achieve acceptable prototype performances. It turns out that our results strongly suggest that commercially available MLCs (even with a non-optimized geometry or material) can already reach this level of performance regarding their heat exchange per unit area.

## Conclusion

The main result of this work allows confirming experimentally the full potential of MLCs with regards efficient exchange heat. Indeed commercially available MLCs are believed to be compatible with prototypes working with heat exchange coefficient h higher than 3400 $W.m^{-2}.K^{-1}$. This value has been extracted for Newtonian case #4 and it is worth noting that non-Newtonian case #2 should have an even higher value of h This study also reveals some design rules that should be followed when it comes to prototyping. First of all, we have to pay attention to the unavoidable electrical connections. Indeed we show that especially the use of solder has to be privileged to improve exchange heat. Consequently any integration into a given prototype has to minimise heat leakage through electrical wires. This favoured path can be seen as a constraint. However, it can also be considered as an opportunity to come up with new designs in order to take advantage of this large heat exchange. Further experiments have to be carried out in order to determine what could be the best way to exchange heat with MLCs. Cooling device prototypes are most of the time fluid-mediated. Therefore, forced convection transfer has to be investigated, in order to determine if it is possible to reach a sufficiently high exchange coefficient to be able to take advantage of terminals properties. We believe that our findings will trigger new investigations in order to improve this crucial heat exchange coefficient.

## Acknowledgments section

This work was entirely supported by the Fonds National de la Recherche (FNR) of Luxembourg through the PEARL "CO-FERMAT" project (P12/4853155) and the InterMobility "MULTICALOR" project (16/1159210).

## List of references